\documentclass[preprint,12pt]{elsarticle}

\usepackage{graphicx}
\usepackage{amssymb}
\usepackage{amsmath}
\usepackage{float}
\usepackage{lineno}

\journal{Annals of Physics}

\begin{document}

\begin{frontmatter}


\title{Switching of the information backflow between a helical spin system and non-Markovian bath}


\author[1]{M. Kaczor}
\author[1]{I.~Tralle}
\author[1]{P.Jakubczyk}
\author[2,3]{S. Stagraczyński}
\author[2]{L. Chotorlishvili\corref{cor1}}
\ead{levan.chotorlishvili@gmail.com}

\address[1]{Institute of Physics, College of Natural Sciences, University of Rzeszów, Pigonia 1, 35-310 Rzeszów, Poland}
\address[2]{Department of Physics and Medical Engineering, Rzeszów University of Technology, 35-959 Rzeszów, Poland}
\address[3]{Institute of Spintronics and Quantum Information, Faculty of Physics, Adam Mickiewicz University, 61-614 Poznań, Poland }

\cortext[cor1]{Corresponding author}
\begin{abstract}
The dissipative dynamics of the spin chain coupled to the non-Markovian magnonic reservoir was studied. The chirality of the chain is formed due to the magnetoelectric coupling. We explored the sign of the trace distance derivative and found the alternating positive/negative periods in system's time evolution. The negative sign is associated with the flow of information from the system to the bath and decrease in states distinguishability, while the positive sign is related to the flow of the information in the opposite direction and increase in distinguishability. We found the distinct effect of the applied electric and magnetic fields. While the Dzyaloshinskii-Moriya interaction and external electric field lead to reshuffling of the periods, the applied magnetic field leads to the swift positive-negative transitions. Thus, in the helical quantum rings coupled to the non-Markovian magnonic baths, it is possible to control the directions of information flow through the external fields.
\end{abstract}

\begin{keyword}
Open Quantum Systems \sep Helical Spin Chains \sep Non-Markovianity \sep Quantum Information

\end{keyword}

\end{frontmatter}


\section{Introduction}
The quantum object in question can be formally split into two parts. Such procedure allows for describing the smaller piece termed as "system" $\hat{H_0}$ through the finite set of canonical quantum operators. At the same time, we postulate the statistical features on the major part termed as a "bath" $\hat{H_b}$. Usually, the expectation values of the bath operators are described by Bose distribution function $n_b=\langle\langle b^+_kb_k\rangle\rangle$. This concept lies at the heart of the open quantum systems theory \cite{diehl2008quantum,campisi2009fluctuation,zhang2012general,
deffner2011nonequilibrium,sieberer2016keldysh,laine2012nonlocal,
rivas2020strong,song2019non,xu2019many,tamascelli2018nonperturbative,
carrega2016energy} and in the present work we follow this standard scheme.

The paradigmatic example of an open quantum system is a two-level system coupled to the phononic (photonic) bath, i.e. the spin-boson model. The coupling to the bath induces decoherence and dissipation effects in the system.  The stochastic thermostat comprises a set of quasiparticle excitations and, in many cases, can possess different physical features.  For example, in the case of spin-phonon interaction, the phononic thermostat relaxes swiftly compared to the spin system. Therefore, the phononic bath recovers instantly from the spin system's impact under the continuous flow of information from the spin system.

The different relaxation time scales between the system and bath permit to exploit the Markov approximation. However, in several cases, the reservoir has the same relaxation time scale. For example, the  NV centers, are surrounded by a nuclear spin bath
\cite{singh2020generation,rabl2009strong,chotorlishvili2013entanglement,mishra2014three}. In this case, the stochastic thermostat is comprised not of phononic modes, but magnonic modes of the surrounding nuclear spins, which have relaxation rates similar to the system itself.
Due to the absence of significant difference between system and environment time scales, Markovian approximation fails. The backflow of information from the environment towards the system leads to the non-Markovian features \cite{breuer2016colloquium,de2017dynamics}.

The non-Markovian system-environment information exchange was studied in the pioneering work \cite{PhysRevLett.100.180402}, within the framework of the stochastic non-Markovian quantum jump method.  Markovian processes reduce the distinguishability between two states, while non-Markovian processes have the opposite effect. Therefore non-Markovianity can be quantified through the trace distance between two density matrices\cite{PhysRevLett.103.210401,li2019non,li2020non}.

In the present work, we aim to explore non-Markovian processes in the helical quantum systems. Multiferroics (MF) constitute a class of materials simultaneously possessing ferroelectric and magnetic properties\cite{azimi2014helical,cheong2007multiferroics,seki2008correlation, katsura2005spin,menzel2012information,azimi2016pulse,wang2018electric,khomeriki2015creation}.
The single-phase MF couples to the external electric field through the magnetoelectric (ME) term. The mathematical structure of the ME term is formally identical to the  Dzyaloshinskii-Moriya (DM) interaction. Therefore MF inherits features of helical systems \cite{wang2019thermally}. Our central interest concerns experimentally feasible helical 1D chains that can be manufactured and embed on a substrate. Such a chain can be used to transfer quantum information\cite{menzel2012information}. Through the exchange interaction, the helical 1D spin chain interacts with the 2D spin lattices of the substrate, which is the essence of the non-Markovian spin bath. For the spins of the non-Markovian bath, we exploit the 
Holstein-Primakoff transformation and map the spin operators of the reservoir into the bosonic magnon operators. Spins of the helical ring coupled to these magnon modes describe the excitation spectrum of the bath. Before proceeding further, we set the Hamiltonian of the model in question.

\begin{figure}[H]
\centering
	\includegraphics[width=0.4\linewidth]{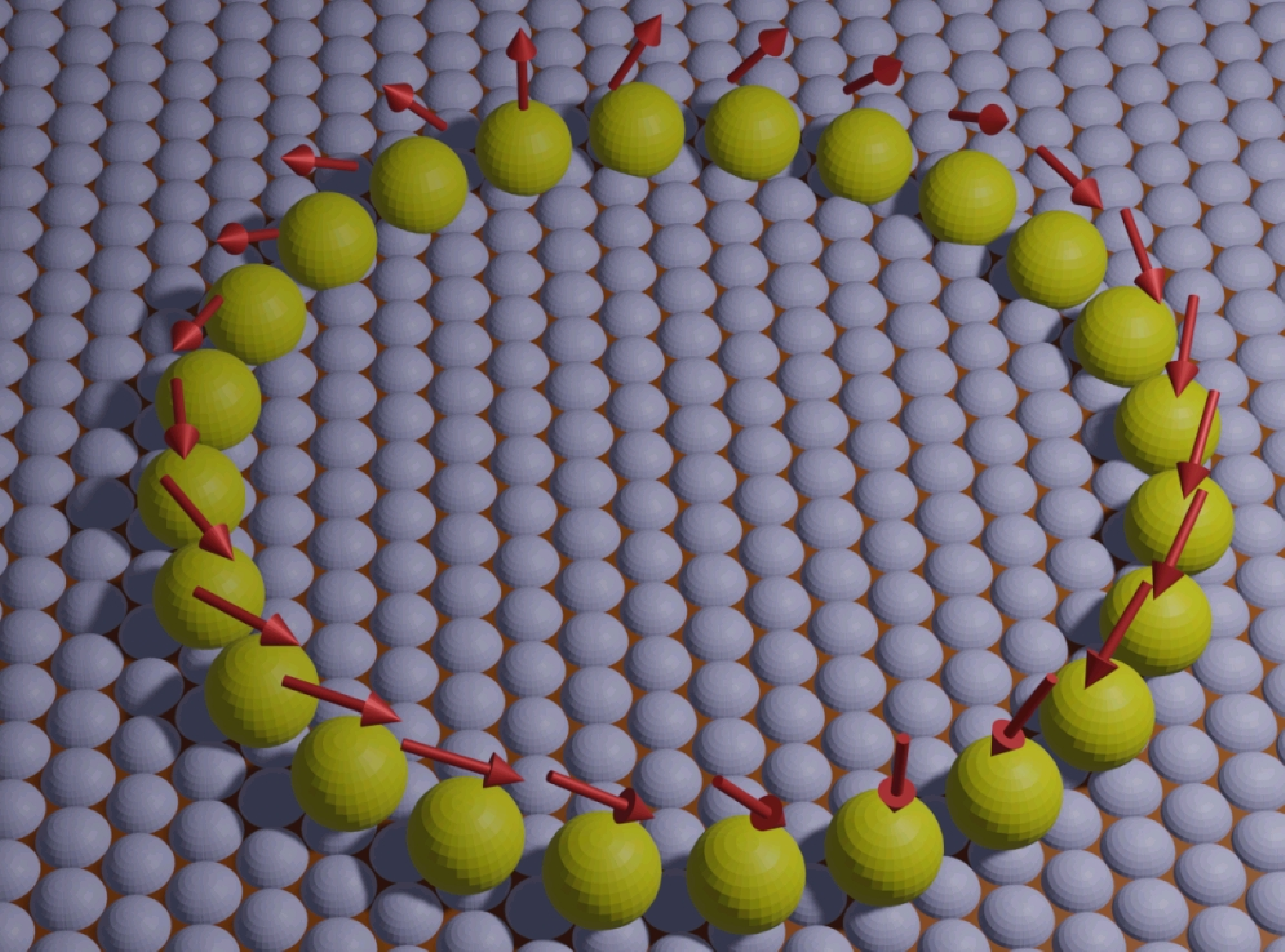}
	\caption{\label{Els} The schematics of the system. The model in question mimics the atomic Fe chains on the $(5\times 1)-Ir(001)$ surface \cite{menzel2012information}. In what follows, we consider the finite length of the chain (up to 50 spins) playing the helical spin system's role. The 2D spin substrate contains many spins and embodies the concept of a non-Markovian spin bath.}
\end{figure}

Let us consider the Hamiltonian of the 1D chiral ring coupled to the non-Markovian spin bath. The total Hamiltonian of the system comprises the Hamiltonian of the chain $\hat{H}_0$, the Hamiltonian of the bath $H_b=\sum\limits_{k=k_1}^{k_{max}}\omega_kb_k^+b_k$, where $k$ innumerates different magnon modes and  $\hat{V}$ is the system-bath interaction term. The total Hamiltonian of the problem reads:
\begin{eqnarray}\label{zerototal Hamiltonian}
&&\hat{H}=\hat{H}_0+\hat{H}_b+\hat{V}.
\end{eqnarray}
The Hamiltonian of the system Fig.(\ref{Els}) has the form:
\begin{eqnarray}\label{secondtotal Hamiltonian}
\hat{H}_0&=&-B\sum\limits_{n=1}^{N}\hat \sigma_n^z+J_1\sum\limits_n\hat{\sigma}_n\hat{\sigma}_{n+1}+J_2\sum\limits_n\hat{\sigma}_n\hat{\sigma}_{n+2}+\nonumber\\&&+D\sum\limits_n\left(\hat{\sigma}_n\times\hat{\sigma}_{n+1}\right)_z.
\end{eqnarray}

The first two terms in Eq. (\ref{secondtotal Hamiltonian}) describe the exchange interaction between nearest and next-nearest spins, the second term represents the DM interaction, which is characterized by the ME coupling constant $D=c_{ME}E$. $\textbf{E}=(0, E,0)$ is the applied external electric field.
By changing the direction of the electric field, we can invert the sign of DM constant. The Hamiltonian $\hat H_0$ commutes with the total $z$ projection of the spin $\hat \sigma^z=\sum\limits_{n=1}^N \hat \sigma^z_n$. Therefore Hamiltonian has a block diagonal structure. Each sector corresponding to the particular $\hat \sigma^z=N, N-1, ...,-N$ can be diagonalized separately. In what follows, we study the low-temperature case and limit the discussion to the single excitation sector. For further convenience, we diagonalize Hamiltonian of the ring and present it in spectral decomposition form: $\hat{H_0}=\sum\limits_{n=1}^NE_n\vert n\rangle\langle n\vert$,
where $E_n,~\vert n\rangle$ are the eigenvalues and eigenfunctions respectively. The interaction term we assume in the following form: $V=\sum\limits_{n}^N\left(\vert n\rangle\langle g\vert B+\vert g\rangle\langle n\vert B^\dagger\right)$, where $B=\sum\limits_k g_kb_k$. We explore the impact of the electric and magnetic fields on the backflow of information from the non-Markovian bath to the system. The paper is organized as follows: in Section \textbf{2}, we describe the analytical solution of Schrödinger equation with total Hamiltonian of the problem. In Section \textbf{3}, we present results of numerical calculation describing the information flow and discuss the possibility of its control through the electric and magnetic field and conclude the work.

\section{Analytical solution}
In what follows, we are interested in the low excitation properties of the system.
We define the zero excitation state of the system $\hat{H}_0$ as $\vert g\rangle=\vert 0_10_2...0_N\rangle$, with energy $E_g=-BN$, for $J_1=-J<0,~J_2=-J_1=J$ and following the standard procedure tackle non-Markovian problem in the single excitation basis
\cite{lorenzo2013tuning}: $\vert n\rangle=\frac{1}{\sqrt{N}}\sum\limits_{l=1}^N\exp(-inl)\vert 1_l\rangle$, where $\vert 1_l\rangle=\vert 0...1_l...0_N\rangle$,
and $E_n=J_1\cos(2\pi n/N)+J_2\cos(4\pi n/N)+D\sin(2\pi n/N)-B(N-1)$. The general state of the coupled system we assumed as follows:
\begin{equation}\label{ansatz}
\vert\phi(t)\rangle=\left(c_0\vert g\rangle+\sum\limits_n^Nc_n(t)\vert n\rangle_S\right) \vert0\rangle_R+\sum_k f_k(t)\vert g\rangle_S\vert 1_k\rangle_R,
\end{equation}
where $\vert 1_k\rangle_R=\vert 1_{{k_1}}\rangle\vert 1_{{k_2}}\rangle...\vert 1_{k_{max}}\rangle$ mean different magnon modes $\omega_k$.
In what follows we consider up to $k_{max}=200$ magnons in the reservoir.
The Schrödinger equation reads:
\begin{eqnarray}\label{the interaction picture}
&&i\frac{d\vert\phi(t)\rangle}{dt}=\hat{H}(t)\vert\phi(t)\rangle,
\end{eqnarray}
or in the explicit form:
\begin{eqnarray}\label{zerothe explicit form}
&&i\frac{dc_n(t)}{dt}=\omega_n c_n(t)+\sum_k g_kf_k(t),\nonumber\\
&&i\frac{df_k(t)}{dt}=\omega_k f_k(t)+\sum_m g_kc_m(t),\nonumber\\
&& n=(1,...N),~~n\neq g.
\end{eqnarray}
Here we introduce the eigenfrequency of the system $\omega_n=(E_n-BN)/\hbar$.
After performing gauge transformations $c_n\equiv e^{-i\omega_n t}c_n$, $f_k\equiv e^{-i\omega_k t}f_k$ we deduce:
\begin{eqnarray}\label{the explicit form}
&&i\frac{dc_n(t)}{dt}=\sum_k g_ke^{i(\omega_{n}-\omega_{k})t}f_k(t),\nonumber\\
&&i\frac{df_k(t)}{dt}=\sum_m g_ke^{-i(\omega_{m}-\omega_{k})t}c_m(t),\nonumber\\
&& n=(1,...N),~~n\neq g.
\end{eqnarray}
The exact set of equations Eq.(\ref{the explicit form}) can be solved numerically, and we use such a solution in the describtion of a information backflow in a system (see Sec. \textbf{3}). 
While the adapted formalism implies a single magnon exchange between the system and bath \cite{li2010non,singh2020generation,man2014non}, the number of magnons in the bath itself is infinite. Here we present approximated analytical solution, following recipes from \cite{li2010non,singh2020generation,breuer2002theory,man2014non} to solve Eq.(\ref{the explicit form}).
The formal solution of the second equation reads:
\begin{eqnarray}\label{formal solution}
&&f_k(t)=-i\sum_m g_k\int_0^te^{-i(\omega_{m}-\omega_k)t_1}c_m(t_1)dt_1.
\end{eqnarray}
We insert this expression into the first equation in Eq.(\ref{the explicit form})
to obtain the closed equation
\begin{eqnarray}\label{closed equations}
&&\frac{dc_n(t)}{dt}=-\sum_m\int_0^tc_m(t_1)\sum_k g_k^2e^{-i(\omega_{g}+\omega_{m}-\omega_{k})(t-t_1)}dt_1.
\end{eqnarray}
Here for the sake of simplicity we took into account that $|\omega_{m}-\omega_{n}|<|\omega_{m}|,~|\omega_{n}|$ and high frequency osculating terms average to zero. For further simplification we postulate spectral properties of the non-Markovian bath
and rewrite Eq.(\ref{closed equations}) in the form
\begin{eqnarray}\label{closed equations2}
&&i\frac{dc_n(t)}{dt}=-\sum_{m\neq n}\int\limits_0^tdt_1f_m(t-t_1)c_m(t_1),\nonumber\\
&&f_m(t-t_1)=\int\limits d\omega J(\omega)\exp\left[-i(\omega_{m}+\omega)(t-t_1)\right],\nonumber\\
&& J(\omega)=\frac{1}{2\pi}\frac{\gamma_0\lambda^2}{(\omega-\omega_c)^2+\lambda^2}.
\end{eqnarray}
Here $\omega_c$ is the center frequency of the bath, $\lambda$ defines the spectral width of the reservoir, $\gamma_0\sim 1/\tau$ is related to the relaxation rate.
To solve Eq.(\ref{closed equations}) we utilize convolution property of Laplace transform:
\begin{eqnarray}\label{Laplace transform}
&&pc_n(p)-c_n(0)=\sum_{m\neq n}c_m(p)f_m(p),\nonumber\\
&&c_n(t)=\frac{1}{i2\pi}\oint\limits_\gamma\exp(pt)c_n(p)dp,
\end{eqnarray}
where contour integration $\gamma$ includes all poles.
The explicit solution is involved and presented in the appendix.

Taking into account the solution we construct the density matrix of the system.
The reduced density matrix reads:
\begin{eqnarray}\label{reduced density matrix}
&& \hat{\varrho}_R^s=\left(1-\sum\limits_n^N\vert c_n(t)\vert^2\right)\vert g\rangle\langle g\vert+\sum_{n=1}^Nc_0c_n^*\vert g\rangle\langle n\vert+\nonumber\\
&& \sum\limits_{n,m}^Nc_n(t)c_m^*(t)\vert n\rangle\langle m\vert+
\sum_{n=1}^Nc_nc_0^*\vert n\rangle\langle g\vert.
\end{eqnarray}
For more details, refer to the appendix.

\section{Control of the backflow of information}

The evolution of the open quantum system can be either Markovian or non-Markovian. Our interest here concerns the case when entire evolution consists of an alternation of Markovian and non-Markovian periods. During the Markovian period, information from the system flows to the bath, and in the non-Markovian case, information flows back to the system. We aim to control the process and switch forward and backflows of information. We associate the direction of the information flow with the sign of the derivative of the trace distance.  The negative sign is associated with the flow of information from the system to the bath. The positive sign is related to the flow of information in the opposite direction. The negative sign of the derivative of trace distance leads to the reduction of distinguishability between the states, while the positive sign enhances it. 

We define the time-derivative of the trace distance as follows: 

\begin{eqnarray}\label{the trace distance as follows}
&& \mathcal{\hat{R}}\left(t,\hat{\rho}_{R,1,2}^s(0)\right)=\frac{d}{dt}\mathcal{D}
\left(\hat{\rho}_{R,1}^s(t),\hat{\rho}_{R,2}^s(t)\right).
\end{eqnarray}
where the trace distance is given by 
\begin{eqnarray}\label{the trace distance explicit}
&& \mathcal{D}
\left(\hat{\rho}_{R,1}^s(t),\hat{\rho}_{R,2}^s(t)\right)=\frac{1}{2}
Tr\left(\vert\hat{\rho}_{R,1}^s(t)-\hat{\rho}_{R,2}^s(t)\vert\right).
\end{eqnarray}
Two density matrices $\hat{\rho}_{R,1}^s(t)$ and $\hat{\rho}_{R,2}^s(t)$ differ in the initial conditions.

 \begin{figure}[H]
 \centering
	\includegraphics[width=0.55\textwidth]{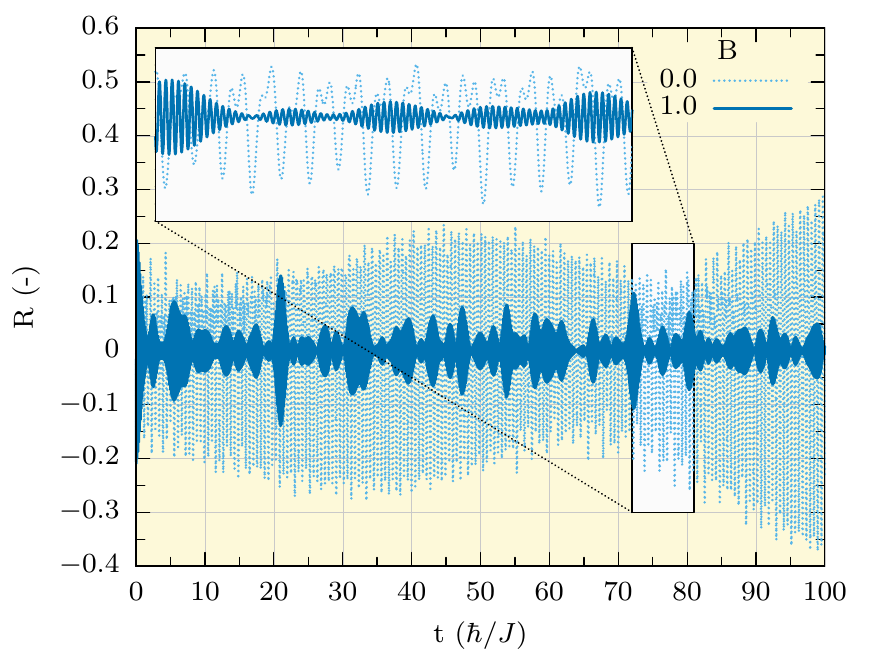}
	\caption{\label{2Els} Time dependence of the derivative of trace distance $R$. Negative parts $R<0$ correspond to the flow of information from the system to the bath, and the positive parts $R>0$ correspond to flow of information in the reverse direction. One can  see that the magnetic field $B$ decreases the modulation amplitude and leads to the fast switching between positive and negative regimes. The values of exchange constants $J=-1$, $J_2=1$, the value of the DM constant $D=0.5$. The number of the spins $N=50$, number of the randomly distributed bath modes $k_{max}=200$. The time scale of the problem is the picosecond order.}
\end{figure}

\begin{figure}[H]
\centering
	\includegraphics[width=0.55\textwidth]{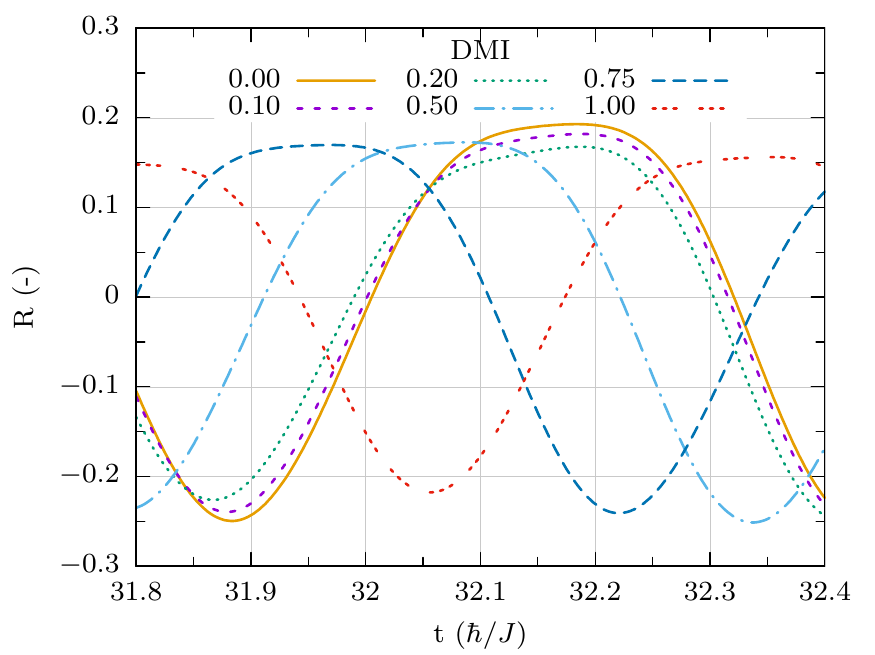}
	\caption{\label{3Els} Time dependence of the derivative of trace distance $R$ without magnetic field $B=0$. Negative parts $R<0$ correspond to the flow of information from the system to the bath, and the positive parts $R>0$ correspond to flow of information in the reverse direction. The values of exchange constants $J=-1$, $J_2=1$. The number of the spins $N=50$, number of the randomly distributed bath modes $k_{max}=200$. The time scale of the problem is the picosecond order.}
\end{figure}

The negative sign in Eq.(\ref{the trace distance as follows}) means that process is Markovian and the positive sign signals the non-Markovianity of the process. We note that the problem of three spins admits the exact analytical solution presented in the appendix.  For the extended system, we carried out numeric calculations that involved 50 spins and 200 modes of the bath. 
The results of numeric calculations are shown in Fig.(\ref{2Els}) and Fig.(\ref{3Els}).
As we see from Fig.(\ref{2Els}), applied external magnetic field leads to the fast 
transitions between positive and negative periods of evolution. The applied external electric field, DM interaction and magnetoelectric coupling leads to the reshuffling of the positive and negative periods -  see Fig.(\ref{3Els}). In particular $D=1$ and $D=0$ cases are shifted by 
$\pi$ in time oscillations. The temporary backflow of information may occur in many physical systems, and it does not always imply strong arguments about thermalization. 
However, negative sign of the derivative of trace distance decreases the distance between states and reduces their distinguishability, while the positive sign increases this feature. Through the external electric field one can control the distinguishability of the states. The effect we noticed is related to the deterministic parameter, such as a magnetic field. The effect we found characterizes the low-temperature properties of helical systems with broken inversion symmetry.

\section{Conclusions}

In the present project, we study dissipative dynamics of the helical quantum ring, coupled to the non-Markovian magnonic reservoir. The chiral spin order in the system is formed due to the magnetoelectric coupling and can be controlled through the applied electric field. We explored the sign of the derivative of trace distance and found that the system's time evolution consists of the alternating positive-negative periods.  The negative (positive) sign is associated with the flow (backflow) of information from the system (bath) to the bath (system).  The negative sign of the derivative of trace distance leads to the reduction of distinguishability between the states, while the positive sign enhances it. Our main findings are as follows: applied external magnetic field leads to the swift positive vs. negative transitions, see Fig. (\ref{2Els}), while the DM interaction and external electric field lead to the reshuffling of the  positive and negative periods, see  Fig.(\ref{3Els}).

\section{Acknowledgment}
This work is supported by the Grant No. FR-19-4049 from Shota Rustaveli National Science Foundation of
Georgia. S.S. acknowledges support from the Norwegian Financial Mechanism under the Polish-Norwegian Research Project NCN GRIEG “2Dtronics,” Project No. 2019/34/H/ST3/00515 

\appendix
\section{Lindblad equation}

From Eq.(\ref{zerothe explicit form}) we deduce the explicit form of the normalization condition:
\begin{eqnarray}\label{the normalization condition simple}
&&\frac{d}{dt}\left\lbrace \sum\limits_n\vert c_n(t)\vert^2+
\sum\limits_k\vert f_k(t)\vert^2\right\rbrace=0,\nonumber \\
&&\sum\limits_n\vert c_n(t)\vert^2+
\sum\limits_k\vert f_k(t)\vert^2 =1-c_0^2.
\end{eqnarray}

Integral in Eq.(\ref{closed equations2}) in the explicit form reads:
\begin{eqnarray}\label{Laplace transform2}
&& f\left(t-t_1\right)=\frac{\gamma_0\lambda^2}{2\pi}\exp\left[-i\left(\omega_m+\omega_c\right)\left(t-t_1\right)\right]\int\limits_{-\infty}^{\infty}\frac{\exp\left[-i(t-t_1)\Omega\right]}{\Omega^2+\lambda^2}d\Omega,\nonumber\\
&&\Omega=\omega-\omega_c .
\end{eqnarray}
After contour integration, from Eq.(\ref{Laplace transform2}) we deduce the following expression:
\begin{eqnarray}\label{Laplace transform3}
 f(t-t_1)&=&-\frac{\gamma_0\lambda}{2}\left\lbrace\exp\left[-(i(\omega_m+\omega_c)+\lambda)(t-t_1)\right]\theta(t-t_1)+\right.\nonumber\\&&\left.+\exp\left[-(i(\omega_m+\omega_c)+\lambda)(t_1-t)\right]\theta(t_1-t)\right\rbrace,
\end{eqnarray}
where $\theta(t-t_1)$ is the Heaviside step function.
For solving Eq.(\ref{Laplace transform}) we exploit the Laplace transform.
The Laplace transform of the function (Eq.(\ref{Laplace transform3})) reads:
\begin{eqnarray}\label{Laplace transform4}
f_m(p)=-\frac{\gamma_0\lambda}{2}\frac{1}{p+\lambda+i(\omega_m+\omega_c)}.
\end{eqnarray}
We present analytic solution Eq.(\ref{Laplace transform}) in the case of three spin ring. Solutions for the larger systems are cumbersome:
\begin{eqnarray}\label{matrixsolution}
&& c_1(p)=
\displaystyle{\frac{\begin{bmatrix} c_1(0) & -f_2(p) & -f_3(p)\\
c_2(0) & p & -f_3(p)\\
c_3(0) & -f_2(p) & p\\
\end{bmatrix}} {D(p)}},~~
c_2(p)=
\dfrac{\begin{bmatrix} p & c_1(0) & -f_3(p)\\
-f_1(p) & c_2(0) & -f_3(p)\\
-f_1(p) & c_3(0) & p\\
\end{bmatrix}} {D(p)},~~\nonumber \\
&& ~~~~~~~~~~~~~~~~~~~~~~~~c_3(p)=
\dfrac{\begin{bmatrix} p & -f_2(p) & c_1(0)\\
-f_1(p) & p & c_2(0)\\
-f_1(p) & -f_2(p) & c_3(0)\\
\end{bmatrix}} {D(p)}
\end{eqnarray}
where:
\begin{equation*}
D(p)=\begin{bmatrix} p & -f_2(p) & -f_3(p)\\
-f_1(p) & p & -f_3(p)\\
-f_1(p) & -f_2(p) & p\\
\end{bmatrix}.
\end{equation*}
For inverse Laplace transform we need to find the roots of the equation
\begin{eqnarray}\label{Laplace transform5}
&& D(p)=p^3-p(f_1(p)f_2(p)+f_1(p)f_3(p)+f_2(p)f_3(p))-\nonumber\\
&& 2f_1(p)f_2(p)f_3(p)=0.
\end{eqnarray}
For the sake of simplicity we assume that $\vert c_1(0)\vert^2=1-\vert c_0\vert^2$
and $c_2(0)=c_3(0)=0$.
The expression for  $D(p)$ can be rewritten in the other form:
\begin{eqnarray}\label{Laplace transform6}
&& D(p) = \frac{ABC p^3 -\frac{\gamma_0^2 \lambda^2}{4}(A+B+C) p + \frac{\gamma_0^3 \lambda^3}{4} }{ABC},
\end{eqnarray}
where for brevity we introduce the following notations:
\begin{eqnarray}\label{Laplace transform7}
&& A=p+ \lambda + i (\omega_1 +\omega_c),\nonumber\\
&& B=p+ \lambda + i (\omega_2 +\omega_c),\nonumber\\
&& C \equiv p + \lambda + i (\omega_3 +\omega_c).
\end{eqnarray}
One can notice that the numerator in the expression above is the sextic polynomial. Depending on the values of parameters, all roots can be different or multiple.  At first, we assume that roots are different and write down the following formulae for the coefficients $c_i, i=1,2,3$:
\begin{eqnarray}\label{Laplace transform7}
&& c_1(p)= \frac{\alpha(p,\omega_1, \omega_2, \omega_3)}{(p-p_1)\times(p-p_2)\times...\times(p-p_6)}, \nonumber\\
&& c_2(p)= \frac{\beta(p,\omega_1, \omega_2, \omega_3)}{(p-p_1)\times(p-p_2)\times...\times(p-p_6)}, \nonumber\\
&& c_3(p)= \frac{\delta(p,\omega_1, \omega_2, \omega_3)}{(p-p_1)\times(p-p_2)\times...\times(p-p_6)},
\end{eqnarray}
where the functions $\alpha, \beta, \delta$ introduced for convenience, take the form:
\begin{eqnarray}\label{Laplace transform8}
&&\alpha (p, \omega_1, \omega_2, \omega_3) = c_1(0) \left(ABC p^2 - \left(\frac{\gamma_0 \lambda}{2} \right)^2 -A \right), \nonumber \\
&&\beta (p, \omega_1, \omega_2, \omega_3) = c_1(0) \left(\frac{\gamma_0 \lambda}{2} \left( B \frac{\gamma_0 \lambda}{2} - BC p \right) \right),~~~~ \nonumber\\
&&\delta (p, \omega_1, \omega_2, \omega_3) = c_1(0) \left(\frac{\gamma_0 \lambda}{2} \left( C \frac{\gamma_0 \lambda}{2} - BC p \right) \right).~~~~
\end{eqnarray}

After calculating the corresponding residua at the simple poles of the integrand (see Eq. 10) and some algebra, we deduce the explicit fromulae for the coefficients $c_i, i=1,2,3$:
\begin{eqnarray}\label{Laplace transform9}
c_1(t)&=&\sum\limits_{n=1}^6\prod\limits_{m=1,m\neq n}^5\left(p_n-p_m\right)^{-1}\times
\sqrt{1-c_0^2}\left((p_n +\lambda + i (\omega_1 +\omega_c))\right. \nonumber\\&& \left. (p_n+ \lambda + i (\omega_2 +\omega_c))(p_n+ \lambda + i (\omega_3 +\omega_c))p_n ^2-\frac{\gamma_0^2 \lambda^2}{4}\right)\exp\left(p_n t\right), \nonumber\\
c_2(t)&=&\sum\limits_{n=1}^6\prod\limits_{m=1,m\neq n}^5\left(p_n-p_m\right)^{-1}\times
\sqrt{1-c_0^2}\left(\frac{\gamma_0^2 \lambda^2}{4}( p_n+ \lambda + i (\omega_2 +\omega_c))\right. \nonumber\\&& \left. -\frac{\gamma_0 \lambda}{2} p_n( p_n+ \lambda + i (\omega_2 +\omega_c))( p_n+ \lambda + i (\omega_3 +\omega_c))\right)\exp\left(p_n t\right), \nonumber\\
c_3(t)&=&\sum\limits_{n=1}^6\prod\limits_{m=1,m\neq n}^5\left(p_n-p_m\right)^{-1}\times
\sqrt{1-c_0^2}\left(\frac{\gamma_0^2 \lambda^2}{4}( p_n+ \lambda + i (\omega_3 +\omega_c))\right. \nonumber\\&& \left. -\frac{\gamma_0 \lambda}{2} p_n( p_n+ \lambda + i (\omega_2 +\omega_c))( p_n+ \lambda + i (\omega_3 +\omega_c))\right)\exp\left(p_n t\right). \nonumber
\end{eqnarray}
Here $p_n$ are the roots of the expression:
\begin{eqnarray}\label{Laplace transform7}
D(p) = \frac{ABC p^3 -\frac{\gamma_0^2 \lambda^2}{4}(A+B+C) p + \frac{\gamma_0^3 \lambda^3}{4} }{ABC}.
\end{eqnarray}
Now we consider the case of two roots with the multiplicity two and four.
Then doing as previously, calculating the residua at the poles of the order of four and two, we get:
\begin{eqnarray}\label{Laplace transform10}
&&c_1(t) = 4 \sqrt{1-c_0^2} \left( \frac{ABC \cdot p_1^2 - \left( \frac{\gamma_0 \lambda}{2} \right)^2 A}{(p_2-p_1)^5} e^{p_1 t}+\frac{ABC \cdot p_2^2 - \left( \frac{\gamma_0 \lambda}{2} \right)^2 A}{(p_1-p_2)^5} e^{p_2 t} \right), \nonumber \\
&& c_2(t) = 4 \sqrt{1-c_0^2} \left(\frac{ \frac{\gamma_0 \lambda}{2} \left( B \frac{\gamma_0 \lambda}{2} - BC p_1 \right)}{(p_2-p_1)^5} e^{p_1 t} +\frac{ \frac{\gamma_0 \lambda}{2} \left( B \frac{\gamma_0 \lambda}{2} - BC p_2 \right)}{(p_1-p_2)^5} e^{p_2 t} \right), \nonumber \\
&&c_3(t) = 4 \sqrt{1-c_0^2} \left(\frac{ \frac{\gamma_0 \lambda}{2} \left( C \frac{\gamma_0 \lambda}{2} - BC p_1 \right)}{(p_2-p_1)^5} e^{p_1 t}+\frac{ \frac{\gamma_0 \lambda}{2} \left( C \frac{\gamma_0 \lambda}{2} - BC p_2 \right)}{(p_1-p_2)^5} e^{p_2 t} \right), \nonumber
\end{eqnarray}
where $A,B,C$ defined as previous and $p_1, p_2$ are the corresponding polynomial roots.
Finally consider the third possibility when the polynomial has three roots, all of the degree 2.
Then doing in the same way, one can get the following expression for the coefficient $c_1$:
\begin{eqnarray}\label{Laplace transform11}
c_1(t) = \frac{2 \alpha(p_1,\omega_1,\omega_2, \omega_3) e^{p_1 t}}{(p_1-p_2)^3(p_1-p_3)^2 + (p_1-p_2)^2(p_1-p_3)^3} +\nonumber\\
\frac{2 \alpha(p_2,\omega_1,\omega_2, \omega_3) e^{p_2 t}}{(p_2- p_1)^3 (p_2 - p_3)^2 + (p_2 - p_1)^2(p_1 - p_3)^3} +~~\nonumber\\
\frac{2 \alpha(p_3,\omega_1,\omega_2, \omega_3) e^{p_3 t}}{(p_3- p_1)^3 (p_3 - p_2)^2 + (p_3 - p_1)^2(p_3 - p_2)^3}.~~~~~~
\end{eqnarray}
The other two coefficients, $c_2(t), c_3(t)$ can be obtained from the previous formula substituting $\alpha(p_i,\omega_1,\omega_2, \omega_3)$
by corresponding values of the functions $\beta(...)$ and $\delta(...)$.

Taking into account normalization condition:
\begin{eqnarray}\label{normalization condition}
\vert c_0(0)\vert^2+\sum\limits_n\vert c_n(t)\vert^2+\sum\limits_k \vert f_k(t)\vert^2=1,
\end{eqnarray}
we obtain the explicit expression of the reduced density matrix of the system:

\begin{eqnarray}
\hat{\varrho}_R^s=\left(1-\sum\limits_n^N\vert c_n(t)\vert^2\right)\vert g\rangle\langle g\vert+\sum_{n=1}^Nc_0c_n^*\vert g\rangle\langle n\vert+\nonumber\\
\sum\limits_{n,m}^Nc_n(t)c_m^*(t)\vert n\rangle\langle m\vert+
\sum_{n=1}^Nc_nc_0^*\vert n\rangle\langle g\vert. ~~~~
\end{eqnarray}
In the matrix form $N=3$:

\begin{eqnarray}\label{In the matrix form}
\hat{\varrho}_R^s=
\begin{bmatrix} (1-\sum_{n=1}^N\vert c_n(t)\vert^2) & c_0(t)c_1^*(t) & c_0(t)c_2^*(t) & c_0(t)c_3^*(t)\\
c_1(t)c_0^*(t) & \vert c_{11}(t)\vert^2 & c_1(t)c_2^*(t) & c_1(t)c_3^*(t)\\
c_2(t)c_0^*(t) & c_2(t)c_1^*(t) & \vert c_{22}(t)\vert^2 & c_2(t)c_3^*(t)\\
c_3(t)c_0^*(t) & c_3(t)c_1^*(t) & c_3(t)c_2^*(t) & \vert c_{33}(t)\vert^2\\
\end{bmatrix},
\end{eqnarray}
where time dependent coefficients are defined in Eq.(\ref{Laplace transform6}).


\bibliographystyle{elsarticle-num-names}
\bibliography{Non-Markovianity}
\end{document}